\documentclass{PoS}
\usepackage{braket}
\usepackage{amsmath,amssymb,mathrsfs,textcomp,amscd,bbm} 

\newcommand{\be}{\begin{equation}}
\newcommand{\ee}{\end{equation}}
\newcommand{\bea}{\begin{eqnarray}}
\newcommand{\eea}{\end{eqnarray}}
\newcommand{\bi}{\begin{itemize}}
\newcommand{\ei}{\end{itemize}}

\def\mev{{\rm MeV}}
\def\gev{{\rm GeV}}
\def\tev{{\rm TeV}}

\def\ev{\mathrm{e\kern-0.1em V}}
\def\kev{\mathrm{ke\kern-0.1em V}}
\def\mev{\mathrm{Me\kern-0.1em V}}
\def\gev{\mathrm{Ge\kern-0.1em V}}
\def\tev{\mathrm{Te\kern-0.1em V}}

\title{Hypercubic effects in semileptonic decays of heavy mesons, toward $B \to \pi \ell\nu$, with Nf=2+1+1 Twisted fermions}

\ShortTitle{Hypercubic effects in semileptonic decays of heavy mesons, toward $B \to \pi \ell\nu$}

\author{Vittorio Lubicz\\
        Universit\'a Roma Tre \& Istituto Nazionale di Fisica Nucleare, Sezione di Roma Tre, Rome, Italy\\
        E-mail: \email{lubicz@fis.uniroma3.it}}

\author{\speaker{Lorenzo Riggio}\\
        Istituto Nazionale di Fisica Nucleare, Sezione di Roma Tre, Rome, Italy\\
        E-mail: \email{lorenzo.riggio@gmail.com}}

\author{Giorgio Salerno\\
        Universit\'a Roma Tre \& Istituto Nazionale di Fisica Nucleare, Sezione di Roma Tre, Rome, Italy\\
        E-mail: \email{salerno@fis.uniroma3.it}}

\author{Silvano Simula\\
        Istituto Nazionale di Fisica Nucleare, Sezione di Roma Tre, Rome, Italy\\
        E-mail: \email{simula@roma3.infn.it}}

\author{\textbf{for the ETM Collaboration}}

\abstract{We present a preliminary study toward a lattice determination of the vector and scalar form factors of the $B \to \pi \ell \nu$ semileptonic decays.
We compute the form factors relative to the transition between heavy-light pseudoscalar mesons, with masses above the physical D-mass, and the pion. 
We simulate heavy-quark masses in the range $m_c^{phys} < m_h < 2m_c^{phys}$.
Lorentz symmetry breaking due to hypercubic effects is clearly observed in the data, and included in the decomposition of the current matrix elements in terms of additional form factors. 
We discuss the size of this breaking as the parent-meson mass increases.
Our analysis is based on the gauge configurations produced by the European Twisted Mass Collaboration with $N_f = 2 + 1 + 1$ flavors of dynamical quarks at three different values of the lattice spacing and with pion masses as small as $210$ MeV.}

\FullConference{The 36th Annual International Symposium on Lattice Field Theory - LATTICE2018\\
		22-28 July, 2018\\
		Michigan State University, East Lansing, Michigan, USA.}

\begin{document}

\section{Introduction and simulation details}
\label{intro}

In all extensions of the Standard Model the transitions between pseudoscalar (PS) mesons can be parametrized in terms of three form factors, namely $f_+$, $f_0$ and $f_T$. Recently, using the gauge configurations generated by the European Twisted Mass Collaboration (ETMC) with $N_f = 2 + 1 + 1$ dynamical quarks \cite{Baron:2010bv,Baron:2011sf}, we have computed the three form factors relevant for the $D \to \pi(K) \ell \nu$ transitions, namely the vector and scalar form factors $f_+^{D \to \pi(K)}(q^2)$ and $f_0^{D \to \pi(K)}(q^2)$ in Ref.~\cite{Lubicz:2017syv}, and the tensor form factor $f_T^{D \to \pi(K)}(q^2)$ in Ref.~\cite{Lubicz:2018rfs} with $q^2 = (p_D - p_{\pi(K)})^2$. The form factors have been evaluated in the whole range of values of $q^2$ accessible by the experiments, i.e.~from $q^2 = 0$ up to $q^2_{\rm max} = (M_D - M_{\pi(K)})^2$. In this contribution we present the results of an exploratory study toward a lattice determination of the vector and scalar form factors for the $B \to \pi \ell \nu$ decays.

The analysis of Refs.~\cite{Lubicz:2017syv,Lubicz:2018rfs} has highlighted an important issue, which warrants further investigations. We found evidence of a remarkable breaking of Lorentz symmetry due to hypercubic effects for all the form factors and we presented a method to subtract the hypercubic artefacts and to recover the Lorentz-invariant form factors in the continuum limit.
These effects appear to be dependent on the difference between the parent and the child meson masses. We thus expect that they will play an even more important role in the determination of the form factors governing semileptonic $B$-meson decays into lighter mesons. It is therefore crucial to have them under control. For this reason we have started an investigation of the semileptonic transition $D^\prime \to \pi \ell \nu$, where $D^\prime$ is an unphysical heavy-light meson, built with a heavier-than-charm quark.

In our calculations quark momenta are injected on the lattice using non-periodic (twisted) boundary conditions and the matrix elements of the vector  current and the scalar density are determined for many kinematical conditions, in which parent and child mesons are  moving or at rest.

The gauge ensembles and the QCD simulations used in this work are the same adopted in Refs.~\cite{Lubicz:2017syv,Lubicz:2018rfs}, where the reader is referred to for details. 
The simulations have been carried out at three different values of the inverse bare lattice coupling $\beta$, at different lattice volumes (with spacial sizes varying between $\simeq 2$ and $\simeq 3$ fm) and for pion masses ranging from $\simeq 210$ to $ \simeq 450$ MeV~\cite{Carrasco:2014cwa}.

\section{Form factors and hypercubic effects}
\label{formfactors}

Let us introduce the local bare vector and scalar operators, $V_\mu = \bar{c}^\prime \gamma_\mu d$ and $S = \bar{c}^\prime d$, where $c^\prime$ is a heavy quark with mass up to twice the physical charm quark mass determined in Ref.~\cite{Carrasco:2014cwa}. Since in our lattice setup we employ maximally twisted fermions, the vector and scalar operators renormalize multiplicatively~\cite{Frezzotti:2003ni}, i.e.~$\widehat{V}_\mu =  {\cal{Z}}_V\cdot V_\mu$ and $\widehat{S} = {\cal{Z}}_P\cdot S$, where $\widehat{V}_\mu$ and $\widehat{S}$ are the renormalized operators with ${\cal Z}_V$ and ${\cal Z}_P$ being the corresponding renormalization constants.

The matrix elements $\braket{\pi(p_\pi) | \widehat{V}_\mu | D^\prime(p_D^\prime)}\equiv\braket{\widehat{V}_\mu}$ and $\braket{\pi(p_\pi) | {S} | D^\prime(p_D^\prime)}\equiv\braket{{S}}$ can be decomposed into the Lorentz-invariant vector and scalar form factors $f_+(q^2)$ and $f_0(q^2)$ as
\begin{eqnarray}
       \label{eq:vector0L}
       \braket{\widehat{V}_0}  & = & (E_{D^\prime} + E_\pi) f_+(q^2) + (E_{D^\prime} - E_\pi) \frac{M_{D^\prime}^2 - M_\pi^2}{q^2} \left[ f_0(q^2) - f_+(q^2) \right] + 
                                                       {\cal{O}}(a^2) ~ , \\
       \label{eq:vectoriL}                                                                      
       \braket{\widehat{V}_{\rm sp}} & = &  \left\{ (p_{D^\prime} + p_\pi) f_+(q^2) + (p_{D^\prime} - p_\pi) \frac{M_{D^\prime}^2 - M_\pi^2}{q^2} 
                                                                                \left[ f_0(q^2) - f_+(q^2) \right] \right\} + {\cal{O}}(a^2) ~ , \\
       \label{eq:scalarL}
       \braket{S} & = & \frac{M_{D^\prime}^2 - M_\pi^2}{\mu_h - \mu_{ud}} f_0(q^2) + {\cal{O}}(a^2) ~ ,
\end{eqnarray}
where $\mu_h$ ($\mu_{ud}$) is the heavy(light) quark bare mass, $\widehat{V}_0$ is the temporal component of the vector current, while $\widehat{V}_{\rm sp}$ is the average of the three spacial components of $\widehat{V}_\mu$ (we inject democratically distributed 3-momenta). The vector and scalar matrix elements can be extracted from the large time distance behavior of three ratios, $R_\mu$ ($\mu = 0, \rm{sp}$) and $R_S$, which are given by
\begin{eqnarray}
\label{eq:Rmu} 
    R_\mu(t,\vec{p}_{D^\prime}, \vec{p}_\pi) &\equiv& 4\, p_{{D^\prime} \mu}\, p_{\pi \mu}\, \frac{C^{{D^\prime}\pi}_{V_\mu}(t, t^\prime, \vec{p}_{D^\prime}, \vec{p}_\pi) \, 
        C^{\pi {D^\prime}}_{V_\mu}(t, t^\prime, \vec{p}_\pi, \vec{p}_{D^\prime})} {C^{\pi \pi}_{V_\mu}(t, t^\prime, \vec{p}_\pi, \vec{p}_\pi) \, 
        C^{{D^\prime}{D^\prime}}_{V_\mu}(t, t^\prime, \vec{p}_{D^\prime}, \vec{p}_{D^\prime})} ~ _{\overrightarrow{t\gg a, \, (t^\prime-t)\gg a}} ~\lvert\braket{\widehat{V}_\mu} \rvert^2 \,,\\[2mm]
    \label{eq:RS}
    R_S(t,\vec{p}_{D^\prime}, \vec{p}_\pi) &\equiv& 4\, E_{D^\prime}\, E_\pi\,\frac{C^{{D^\prime}\pi}_S(t, t^\prime,\vec{p}_{D^\prime}, \vec{p}_\pi)\, C^{\pi {D^\prime}}_S(t, t^\prime,\vec{p}_\pi, \vec{p}_{D^\prime})} {\widetilde{C}_2^{D^\prime} \left(t^\prime,\vec{p}_{{D^\prime}} \right)\, \widetilde{C}_2^\pi \left(t^\prime,\vec{p}_{\pi} \right)} \,~ _{\overrightarrow{t\gg a, \, (t^\prime-t)\gg a}}~\lvert\braket{S} \rvert^2  \,.
\end{eqnarray}
In Eqs.~(\ref{eq:Rmu})-(\ref{eq:RS}) $C^{D^\prime\pi}_{\Gamma}$ ($\Gamma=V_\mu,\,S$) and $\widetilde{C}_2^M$ are respectively the 3-point correlation function between the $D^\prime$ and the $\pi$ mesons and the 2-point correlation function for the meson $M$ in which the backward signal has been cancelled (see Ref.~\cite{Lubicz:2017syv}). At large time distances they are defined as
\begin{eqnarray}
\label{eq:C3_larget}    
        C^{{D^\prime}\pi}_{\widehat{\Gamma}}\left( t,\, t^\prime,\, \vec{p}_{D^\prime},\, \vec{p}_\pi \right) && ~ _{\overrightarrow{t\gg a\,, \, (t^\prime-t)\gg a}} ~
              Z_\pi Z_{D^\prime}^* \braket{\pi(p_\pi)|\widehat{\Gamma}|{D^\prime}(p_{D^\prime})}\, e^{-E_{D^\prime} t} \, e^{-E_\pi (t^\prime - t)} / (4E_\pi E_{D^\prime}) ~ , \\
\label{eq:C2_tilde_larget}             
        \widetilde{C}_2^{M} \left(t,\,\vec{p}_{M}\right) && ~ _{\overrightarrow{t \gg a}} ~ Z_M \, e^{-E_M t} / (2 E_M) ~ ,
\end{eqnarray}
where $E_{D^\prime}$ and $E_{\pi}$ are the energies of the $D^\prime$ and $\pi$ mesons, while $Z_D^\prime$ and $Z_\pi$ are the matrix elements $\braket{0\lvert\,P_5^{D^
\prime}(0)\,\rvert\,D^\prime(\vec{p}_{D^\prime}) }$ and $\braket{0\lvert\,P_5^\pi(0)\,\rvert\,\pi(\vec{p}_{\pi})}$, which depend on the meson momenta $\vec{p}_D^\prime$ and $\vec{p}_
\pi$ because of the use of smeared interpolating fields.
From the 2- and 3-point correlators we are able to extract the matrix elements $\braket{\widehat{V}_\mu}$ and $\braket{S}$, which allow us to determine $f_+(q^2)$ and $f_0(q^2)$ as the best-fit values of Eqs.~(\ref{eq:vector0L})-(\ref{eq:scalarL}). 

\begin{figure}[htb!]
\centering
\includegraphics[width=6.65cm,clip]{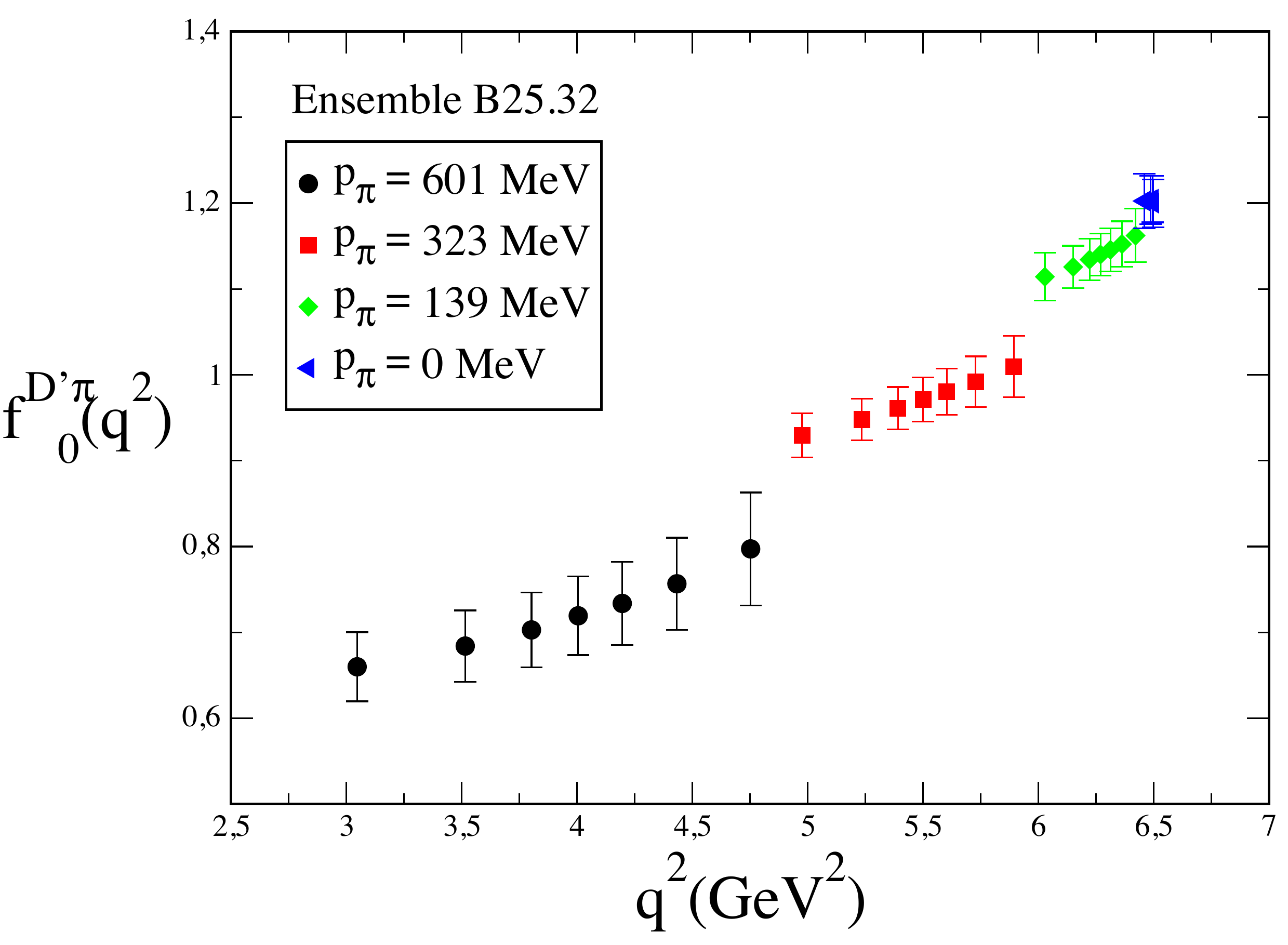}
\includegraphics[width=6.65cm,clip]{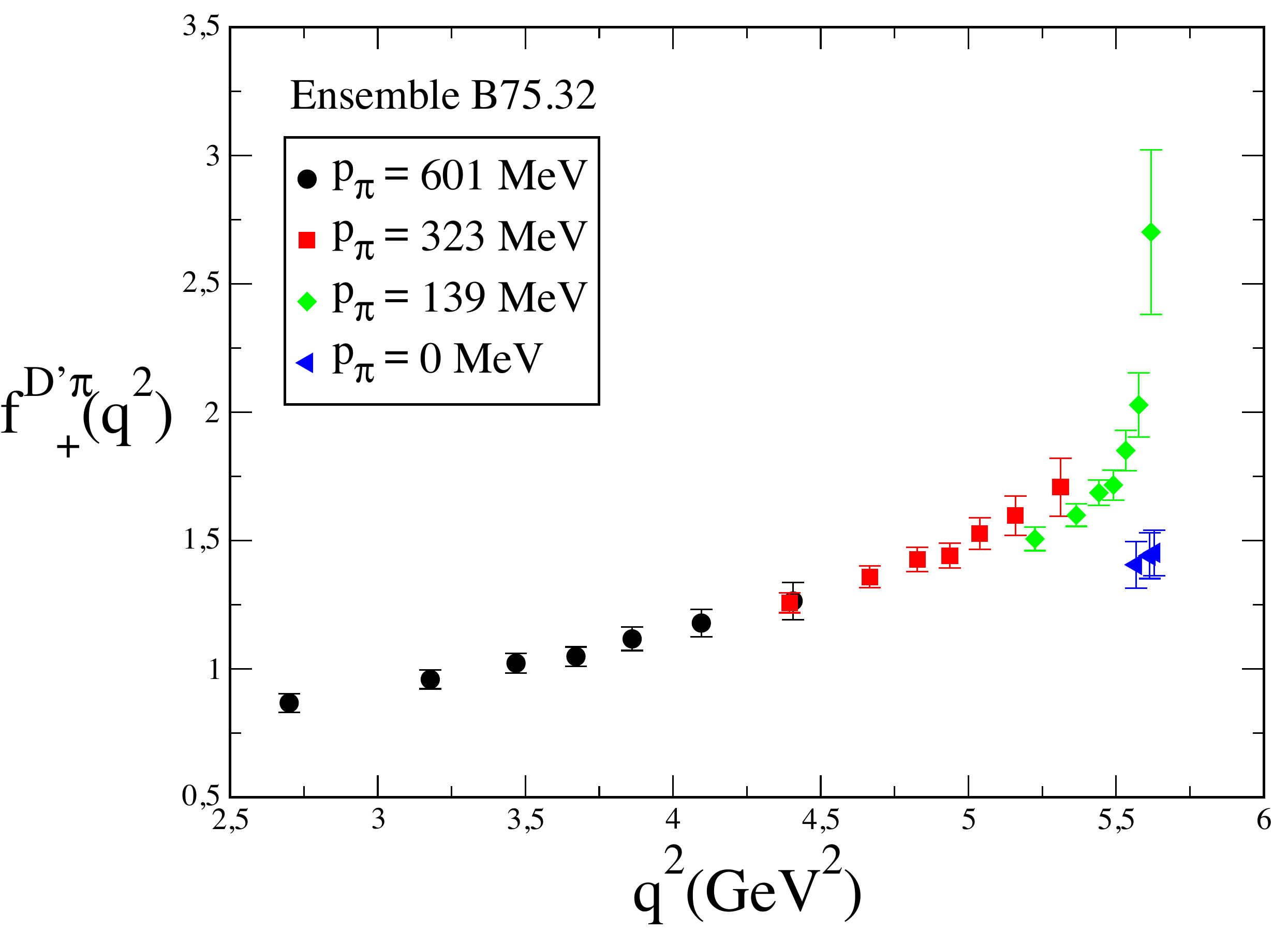}
\vspace{-0.25cm}
\caption{\it \footnotesize Momentum dependence of the scalar form factor for the ensemble B25.32 (left panel) and of the vector form factor for the ensemble B75.32 (right panel), corresponding respectively to $M_\pi \approx 250$ MeV and $M_\pi \approx 400$ MeV~\protect\cite{Carrasco:2014cwa}. In both cases the unphysical $D^\prime$-meson has a mass of $\simeq 2.7$ GeV corresponding to a heavy-quark mass $m_h \simeq 2.13$ GeV. Different markers and colors distinguish different values of the pion momentum.}
\label{fig:fishbone}
\end{figure}

The momentum dependencies of the vector and scalar form factors are illustrated in Fig.~\ref{fig:fishbone}, where different markers and colors correspond to different values of the child meson momentum for two different ensembles. The unphysical $D^\prime$-meson has a mass of $\simeq 2.7$ GeV corresponding to a heavy-quark mass $m_h \simeq 2.13$ GeV.
If the Lorentz-covariant decomposition~(\ref{eq:vector0L})-(\ref{eq:scalarL}) were adequate to describe the lattice data, the extracted form factors would depend only on the squared 4-momentum transfer $q^2$ (and on meson masses).
This is not the case and an extra dependence on the value of the meson 3-momenta is clearly visible in Fig.~\ref{fig:fishbone}.
The origin of this effect is that the lattice breaks Lorentz symmetry and it is invariant only under discrete hypercubic rotations. Therefore, the form factors determined form Eqs.~(\ref{eq:vector0L})-(\ref{eq:scalarL}) depend also on hypercubic invariants.

\begin{figure}[htb!] 
\centering
\makebox[\textwidth][c]{
\includegraphics[width=6.65cm,clip]{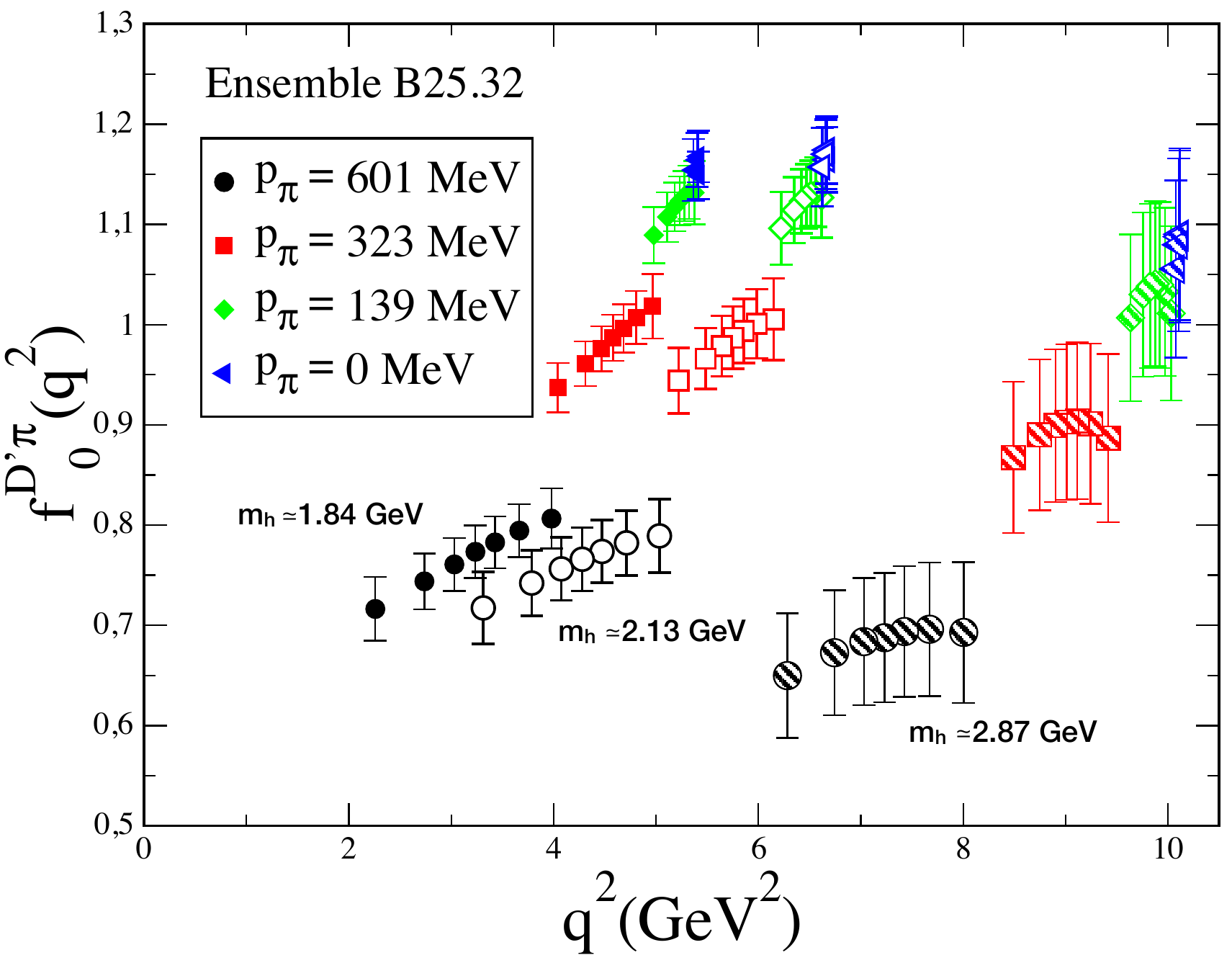}
\includegraphics[width=6.65cm,clip]{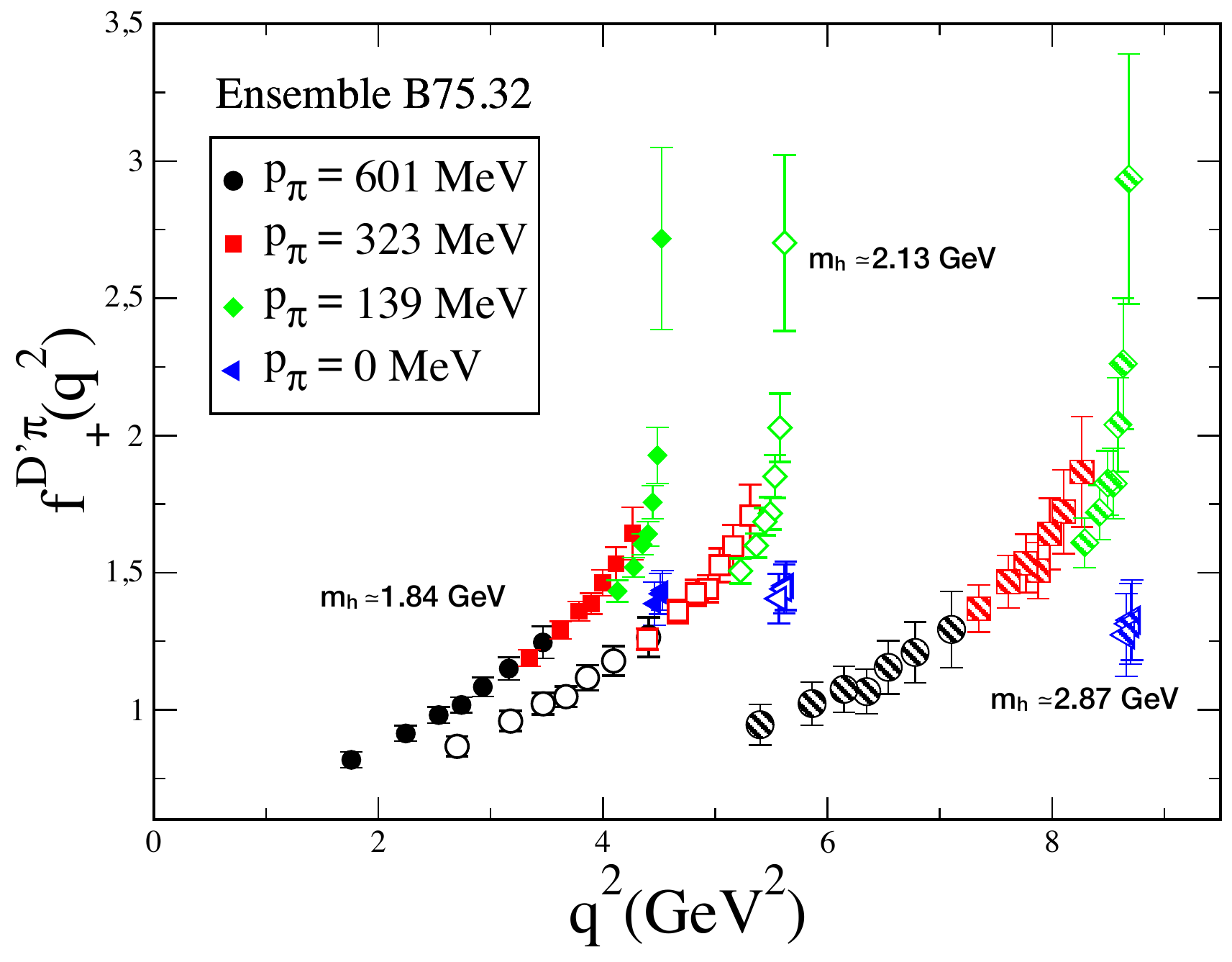}
}
\makebox[\textwidth][c]{
\includegraphics[width=6.65cm,clip]{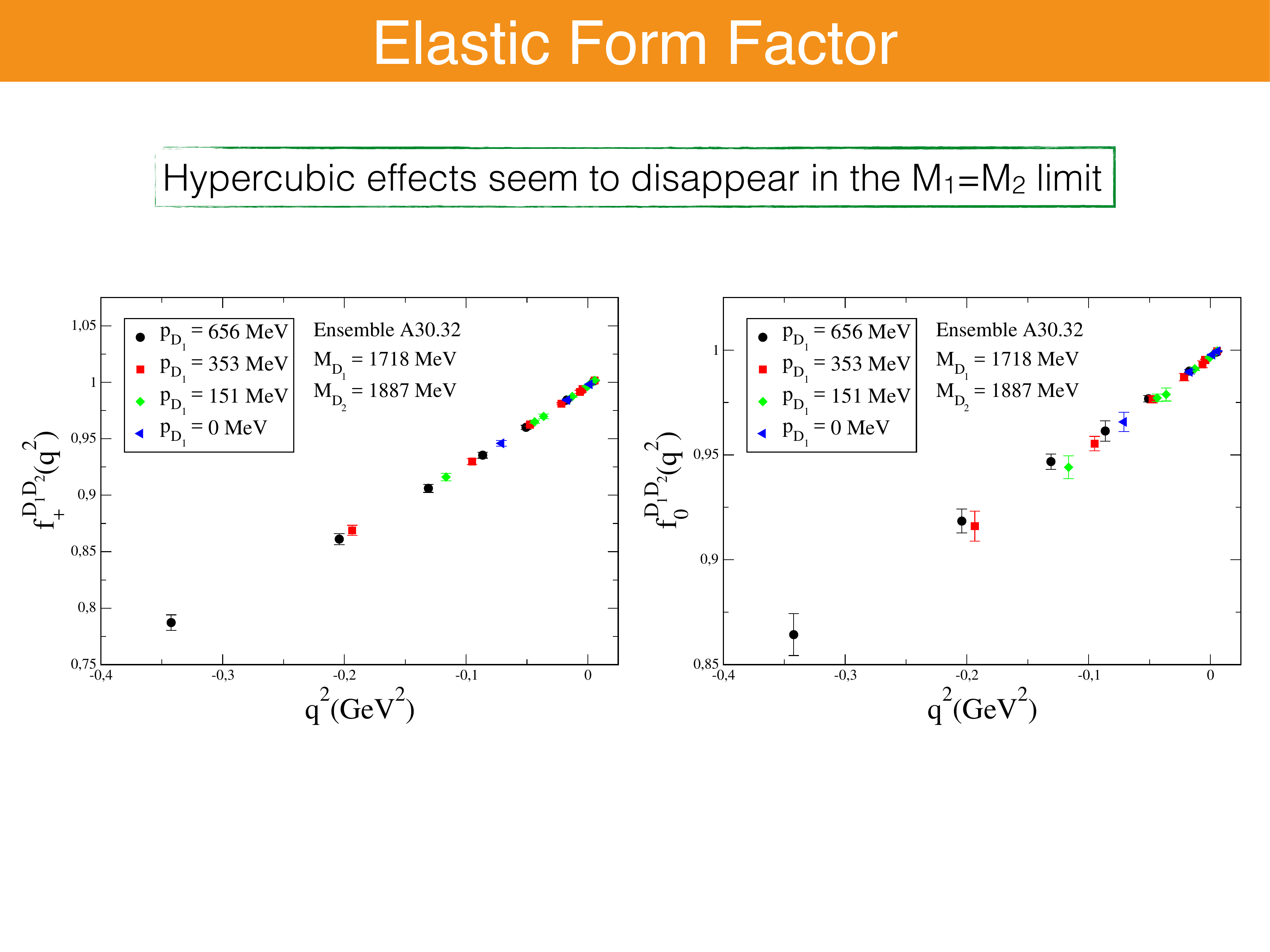}
\includegraphics[width=6.65cm,clip]{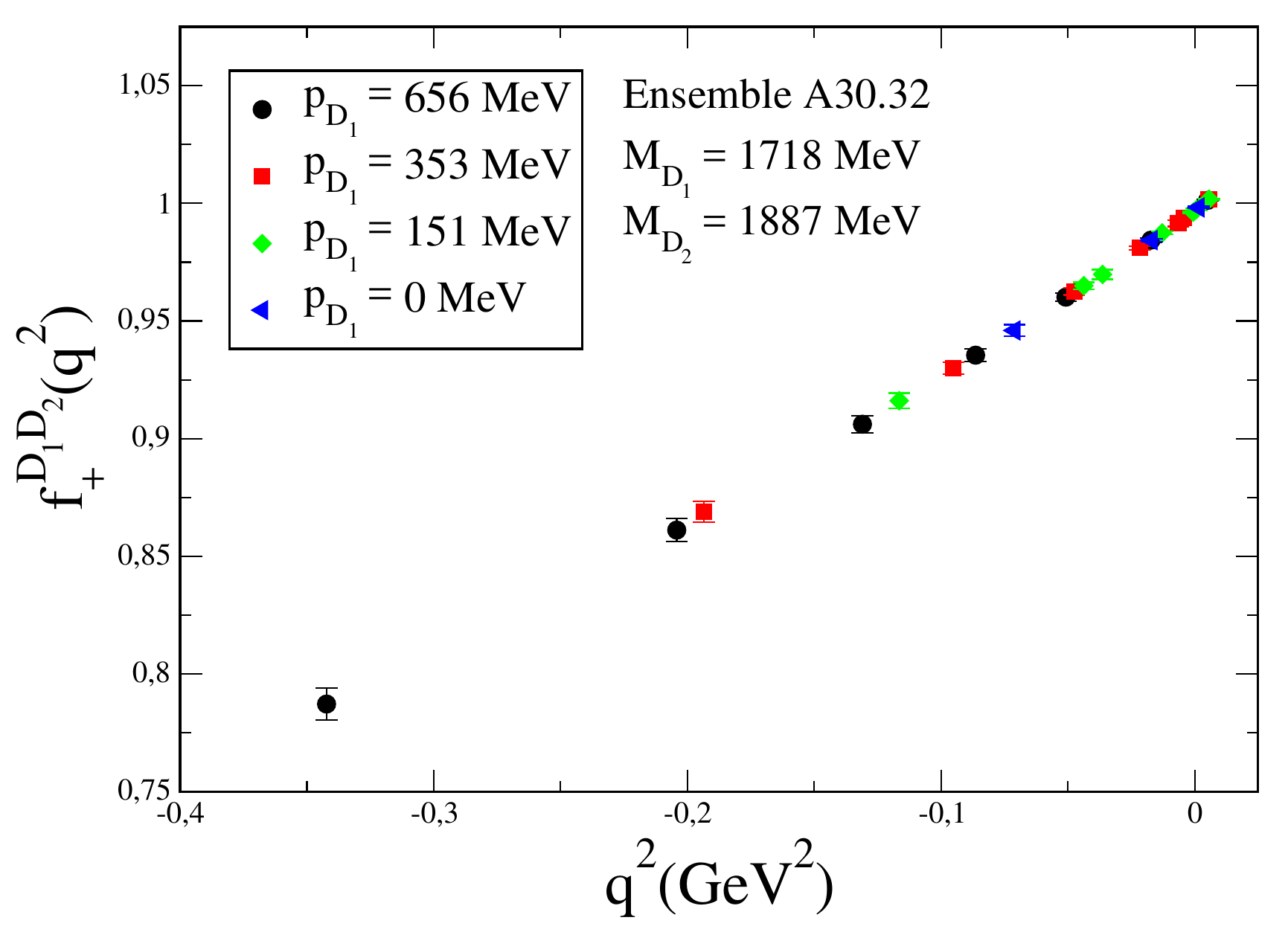}
}
\vspace{-0.75cm}
\caption{\it \footnotesize Upper panels: the vector and scalar form factors for a transition between two non-degenerate PS mesons for three different values of the heavy-quark mass. Lower panels: the vector and scalar form factors for transitions between two nearly degenerate PS mesons close to the physical D-meson.}
\label{fig:elastic_vs_nonelastic}
\end{figure}

The upper panels of Fig.~\ref{fig:elastic_vs_nonelastic} shows how the hypercubic effects on the vector and scalar form factors change with the mass difference ($M_{D^\prime} - M_\pi$). In the lower panels the case of the transition between two nearly degenerate PS mesons close to the physical D-meson is shown. Within the statistical uncertainties the corresponding form factors exhibit no evidence of hypercubic effects.

\section{Global fit}
\label{globalfit}

A possible way to describe hypercubic artifacts affecting the form factors $f_+$ and $f_0$ is to address them directly on the vector and scalar matrix elements (\ref{eq:vector0L})-(\ref{eq:scalarL}). 
The analysis closely follows the strategy presented in Ref.~\cite{Lubicz:2017syv} for the vector and scalar $D \to \pi(K)$ form factors. We refer the interested reader to the more detailed discussion presented in Refs.~\cite{Lubicz:2017syv,Lubicz:2018rfs}.
To illustrate the procedure we consider the following decomposition of the vector matrix elements
\begin{equation}
    \braket{\pi(p_\pi) | \widehat{V}_\mu^E | {D^\prime}(p_{D^\prime})} = \braket{\widehat{V}_\mu^E}_{\rm Lor} + \braket{\widehat{V}_\mu^E}_{\rm hyp} ~ ,
    \label{eq:vector_decomposition}
\end{equation}
where $\braket{\widehat{V}_\mu^E}_{\rm Lor}$ is the Lorentz-covariant term
\begin{equation}
    \braket{\widehat{V}_\mu^E}_{\rm Lor} = P_\mu^E~ f_+(q^2, a^2) + q_\mu^E ~ \frac{M_{D^\prime}^2 - M_{\pi}^2}{q^2} \left[ f_0(q^2, a^2) - f_+(q^2, a^2) \right] ~ ,
    \label{eq:vector_Lorentz}
\end{equation}
while $\braket{\widehat{V}_\mu^E}_{\rm hyp}$ is given by
\begin{equation}
    \braket{\widehat{V}_\mu^E}_{\rm hyp} = a^2 \left[ \left( q_\mu^E \right)^3 ~ H_1+ \left( q_\mu^E \right)^2 P_\mu^E ~ H_2 + 
                                                                    q_\mu^E \left( P_\mu^E \right)^2 ~ H_3 + \left( P_\mu^E \right)^3 ~ H_4 \right] ~ .
    \label{eq:vector_hypercubic}
\end{equation}
The suffix $E$ emphasizes that all the relations are written in the Euclidean space, e.g.~$q_\mu^E = \left( \vec{q}, ~ q_4 \right) = \left( \vec{q}, ~ iq_0 \right)$ so that $\sum_\mu q_\mu^E q_\mu^E = - q^2$.
The new structures, allowed by the lattice hypercubic symmetry, introduce the new form factors $H_i$ ($i = 1, ..., 4$).
For them we have adopted a polynomial expansion $H_i(z) = d_0^i + d_1^i z + d_2^i z^2$ given in terms of the $z$ variable~\cite{Boyd:1995cf}.
The coefficients $d_{0,1,2}^i$, which depend on the light-quark mass, are free parameters in the fitting procedure.

\begin{figure}[htb!]
\centering
\includegraphics[width=6.65cm,clip]{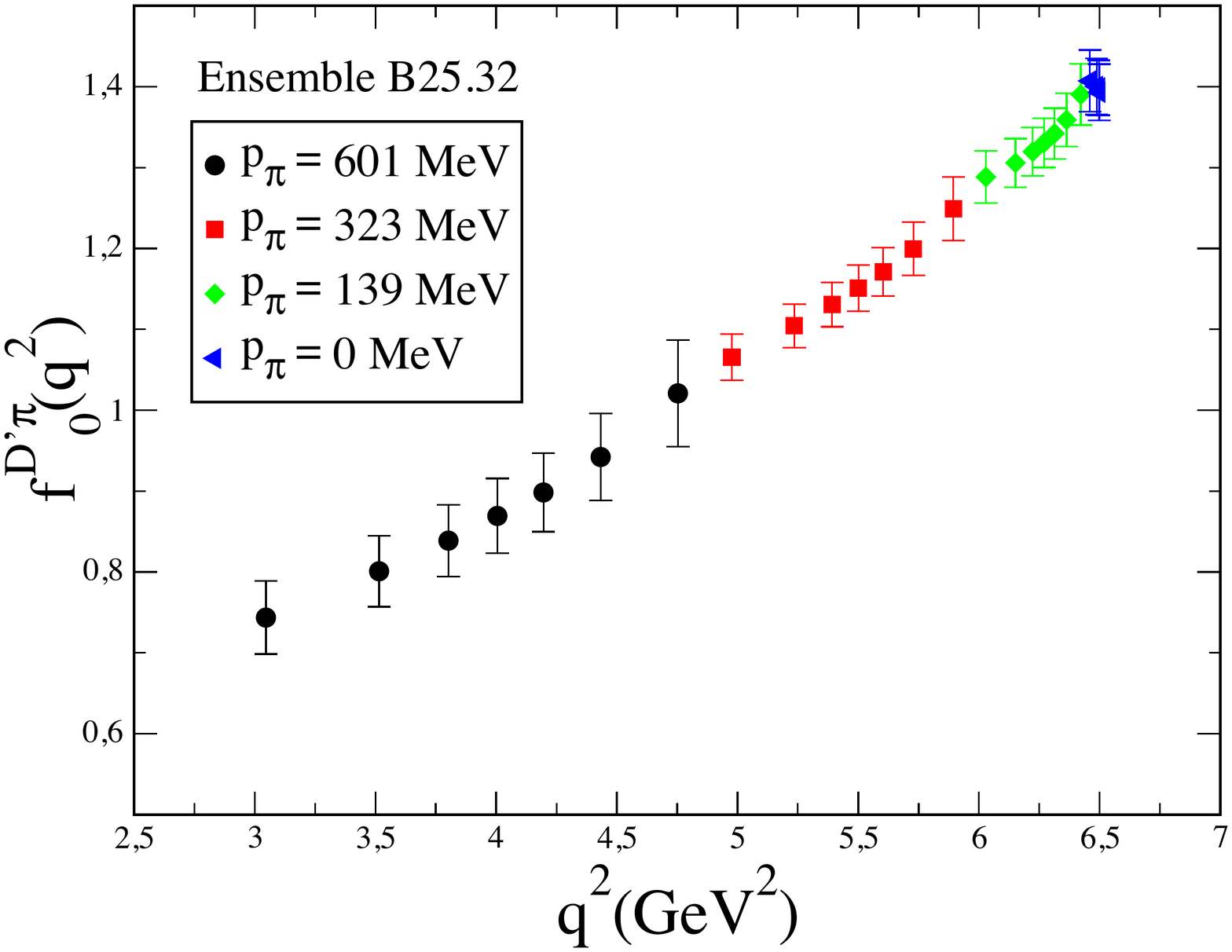}
\includegraphics[width=6.65cm,clip]{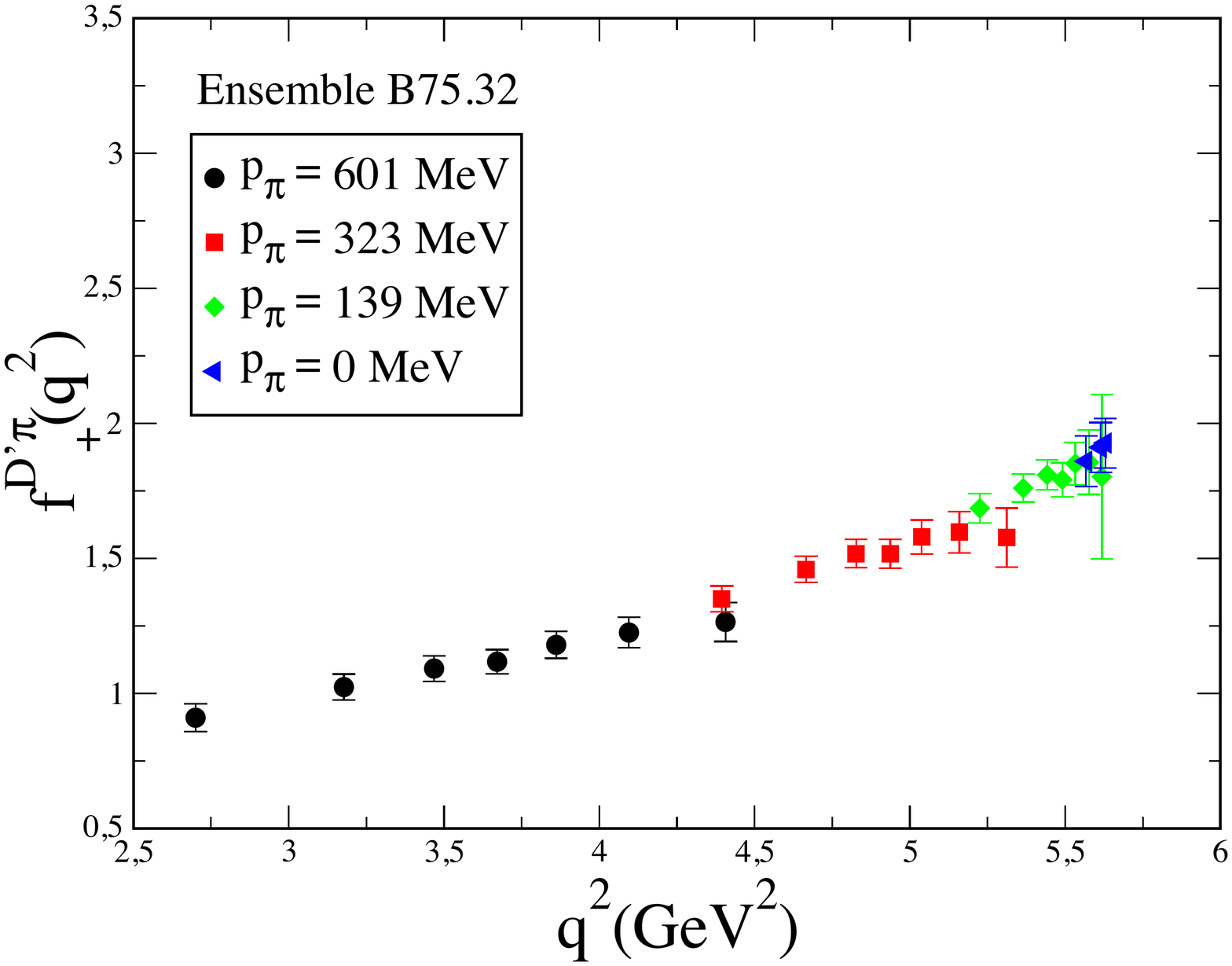}
\vspace{-0.25cm}
\caption{\it \footnotesize The same as in Fig.~\ref{fig:fishbone} after subtracting the hypercubic contributions as determined by the global fit.}
\label{fig:corrected}
\end{figure}

As for the Lorentz-invariant terms $\braket{\widehat{V}_\mu^E}_{\rm Lor}$ and $\braket{S}_{\rm Lor}$, the form factors $f_{+,0}(q^2, a^2)$ can be parametrized by the modified z-expansion of Ref.~\cite{Bourrely:2008za}, viz.
 \begin{eqnarray}
   \label{eq:z-exp_f+}
    f_+^{{D^\prime} \to \pi}(q^2, a^2) & = & \frac{f^{{D^\prime} \to \pi}(0, a^2) + c_+(a^2)\, (z - z_0)
                                                               \left(1 + \frac{z + z_0}{2} \right)}{1 - q^2 \left( 1 + P_+ a^2 \right) / \left( M_{D^\prime}^2 + \Delta^2 \right)} ~ ,\\
   \label{eq:z-exp_f0}
    f_0^{{D^\prime} \to \pi}(q^2, a^2) & = & \frac{f^{{D^\prime} \to \pi}(0, a^2) + c_0(a^2)\, (z - z_0) 
                                                               \left(1 + \frac{z + z_0}{2} \right)}{1 - q^2 / M_S^2} ~ ,
 \end{eqnarray}
where for $c_{+,0}$ we have adopted a linear dependence in $a^2$, $M_{D^\prime}$ is the PS mass calculated on the lattice, $\Delta^2 \equiv M_{D*}^2 - M_D^2$ is taken from experiments and $M_S$ is left as a free parameter.
For the vector form factor at zero 4-momentum transfer, $f^{{D^\prime} \to \pi}(0, a^2)$, we use the following Ansatz
\bea
    \label{eq:ChLim}
    f^{{D^\prime} \to \pi}(0, a^2)\!\!&=&\!\! F_+ \left[ 1 + - \frac{3}{4} \left( 1 + 3 \widehat{g}^2 \right) \xi_\ell \log\xi_\ell + b_1\, \xi_\ell + b_2 \, \xi^2_\ell + b_3 \, a^2 \right] ~,
\eea
where the coefficients $F_+$, $b_1$, $b_2$ and $b_3$ are treated as free parameters.
In Fig.~\ref{fig:corrected} we show the same form factors as in Fig.~\ref{fig:fishbone} after the hypercubic contributions determined by the global fit have been subtracted. It can be seen that the vector and scalar form factors depend now only on the $4-$momentum transfer $q^2$. 

\begin{figure}[htb!] 
\centering
\makebox[\textwidth][c]{
\includegraphics[width=6.65cm,clip]{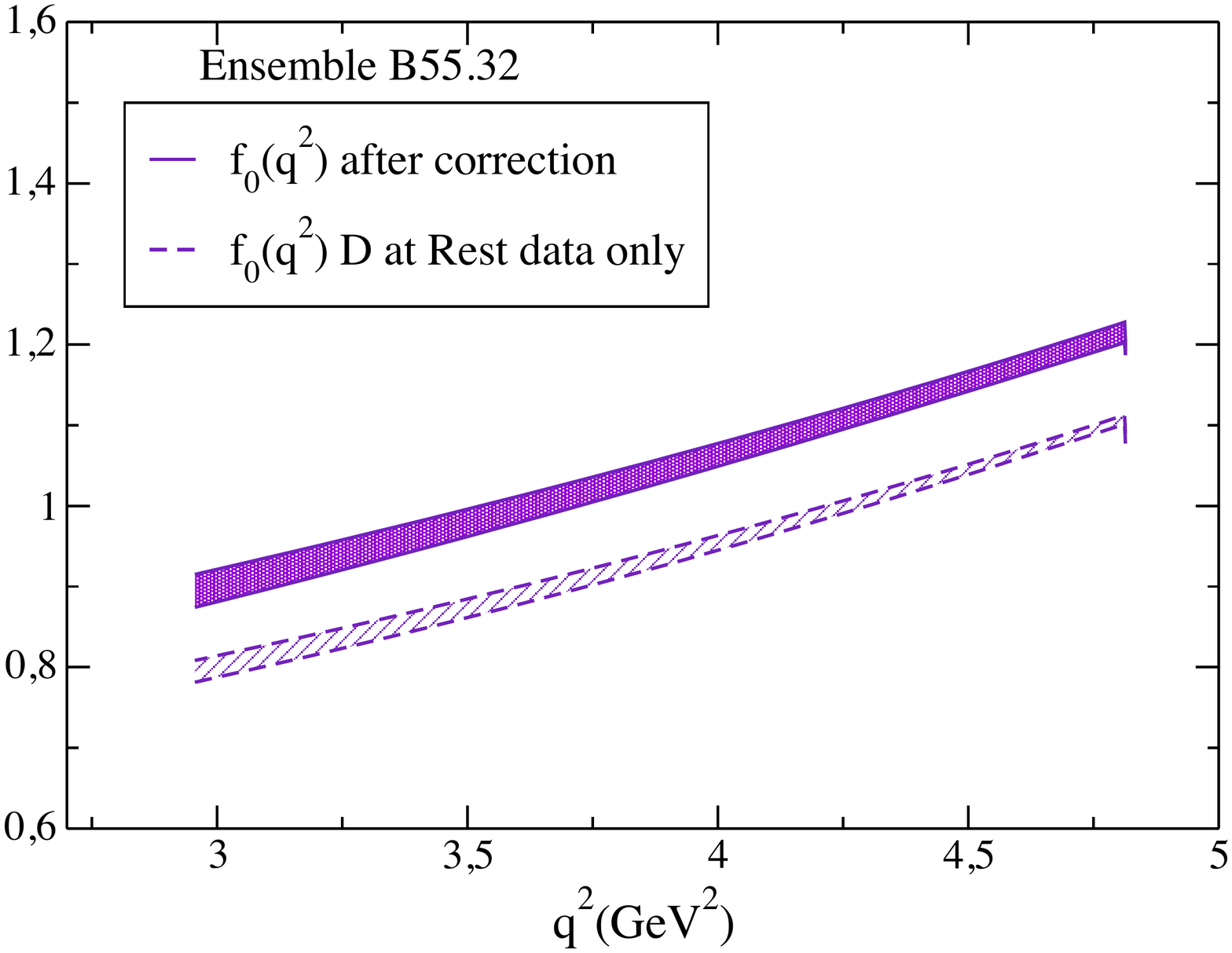}
\includegraphics[width=6.65cm,clip]{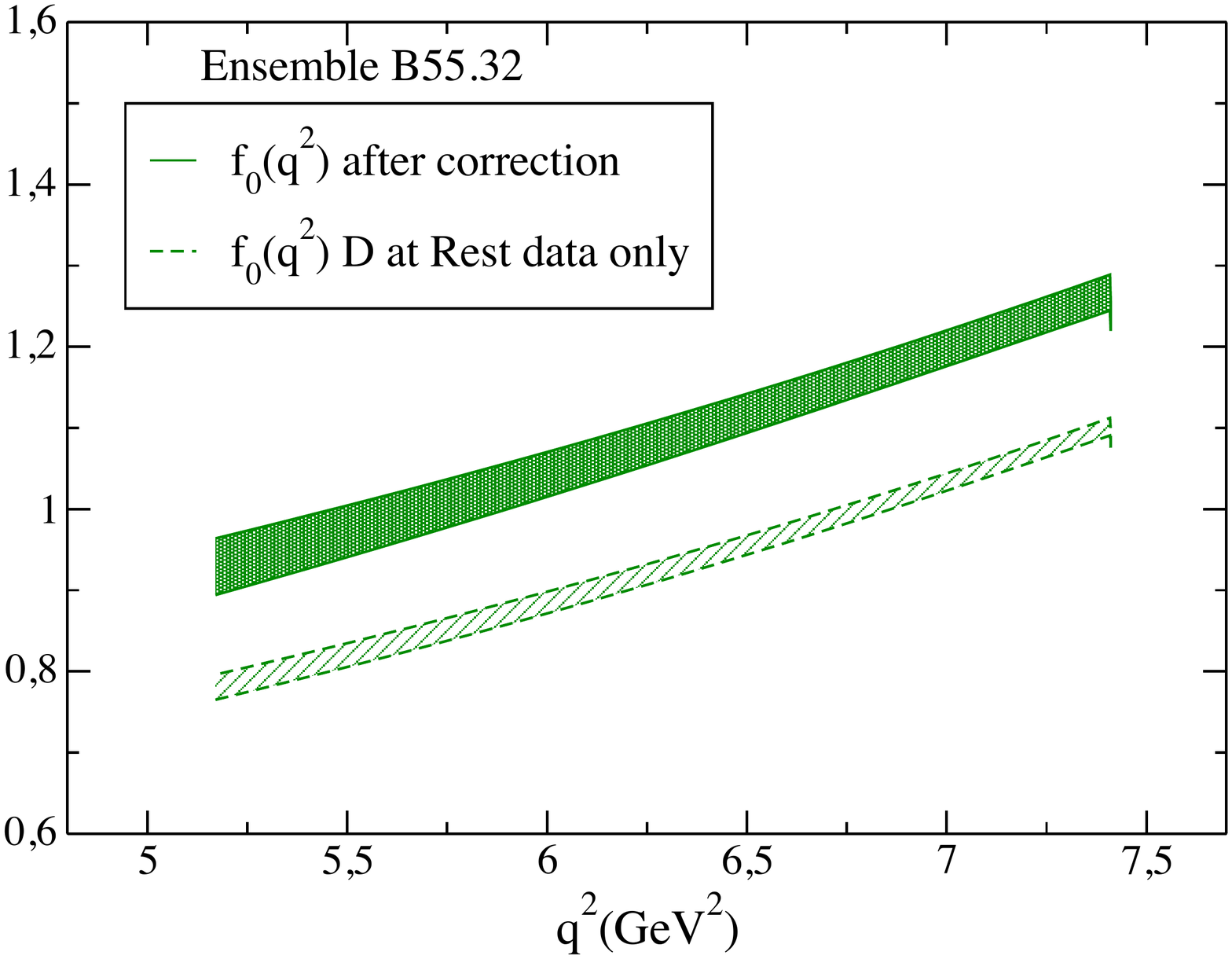}
}
\makebox[\textwidth][c]{
\includegraphics[width=6.65cm,clip]{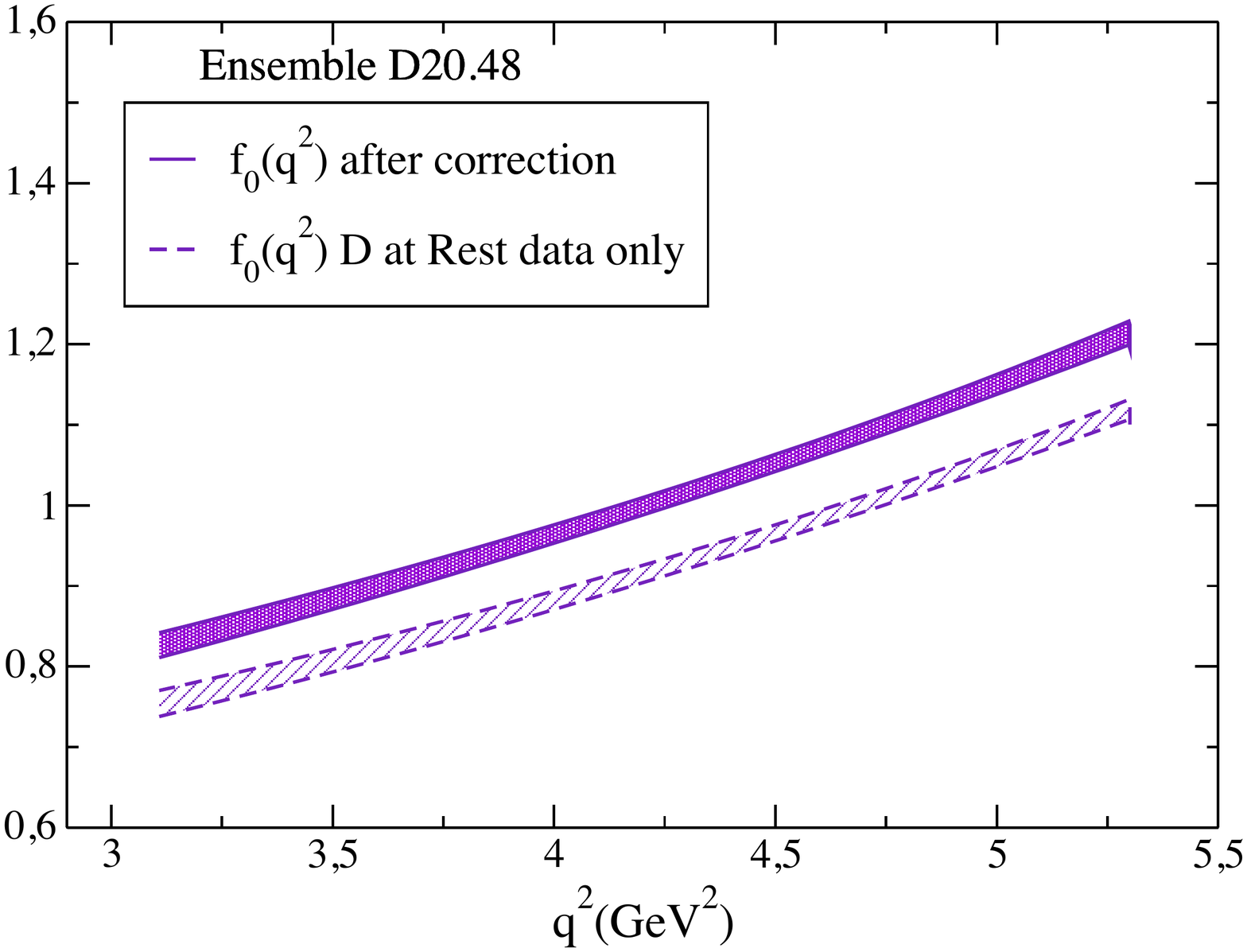}
\includegraphics[width=6.65cm,clip]{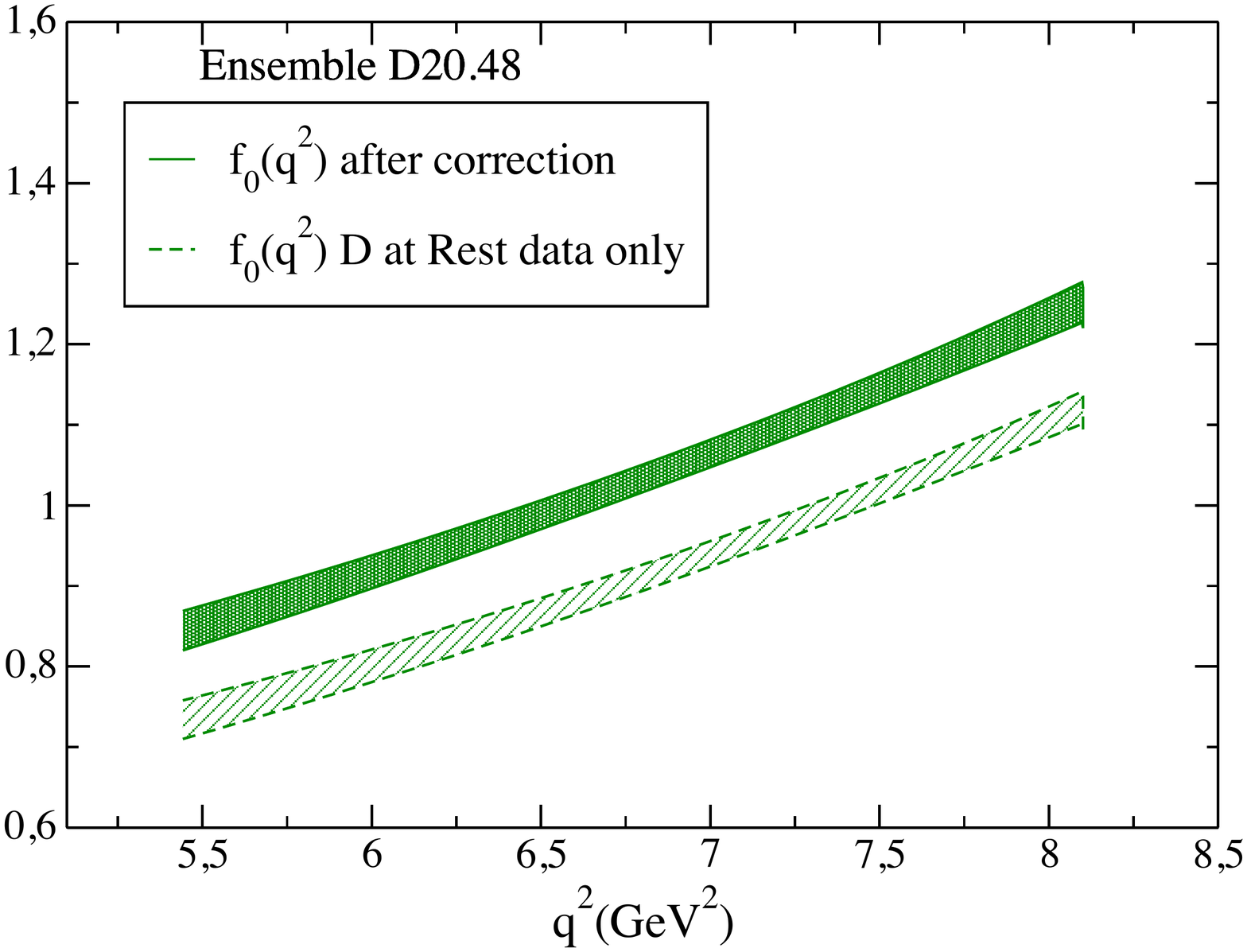}
}
\vspace{-0.75cm}
\caption{\it \footnotesize Momentum dependence of the scalar form factor $f_0(q^2)$ corresponding to $m_h \simeq 1.85$ GeV (left panels) and $m_h \simeq 2.50$ GeV (right panels) in the case of the gauge ensembles B55.32 and D20.48~\protect\cite{Carrasco:2014cwa}. The filled bands correspond to the global fit where all the available data are included after subtraction of the hypercubic effects. The dashed bands correspond to the fit of the data corresponding only to the heavy meson at rest without taking into account hypercubic corrections. The pion mass and the lattice spacing are respectively equal to $M_\pi \simeq 375$ MeV and $a \simeq 0.0815$ fm for the ensemble B55.32, and $M_\pi \simeq 250$ MeV and $a \simeq 0.0620$ fm in the case of the ensemble D20.48.}
\label{fig:HMaRcomparison_latt}
\end{figure}

The impact of the hypercubic effects at fixed lattice spacing is shown in Fig.~\ref{fig:HMaRcomparison_latt}, where we compare the scalar form factor $f_0(q^2)$ obtained by fitting the data from Eqs.~(\ref{eq:vector0L})-(\ref{eq:scalarL}) corresponding only to the heavy meson at rest with the one obtained using all the available data after subtracting the hypercubic effects determined from the global fit.

\begin{figure}[htb!] 
\centering
\makebox[\textwidth][c]{
\includegraphics[width=7.10cm,clip]{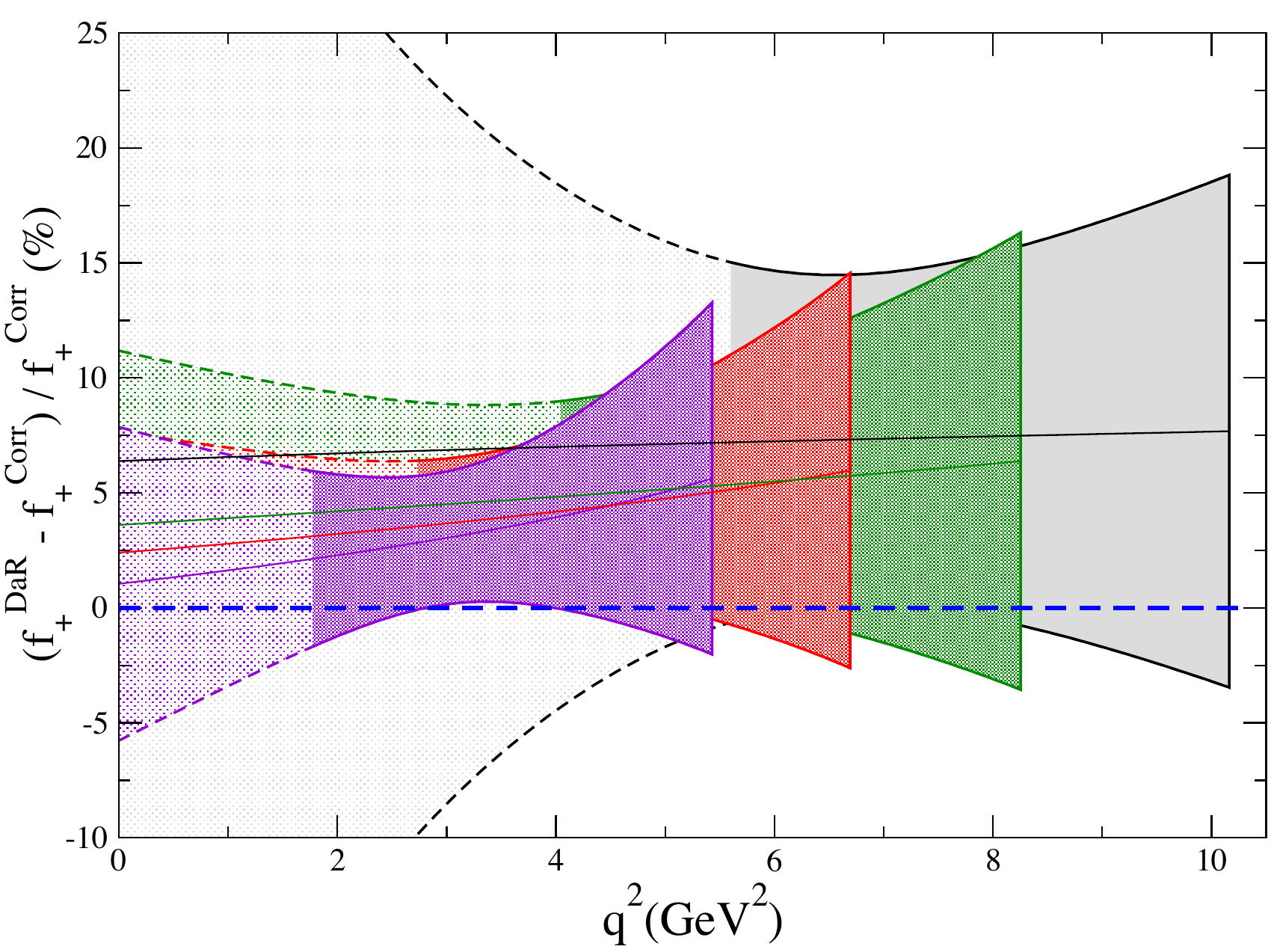}
\includegraphics[width=7.10cm,clip]{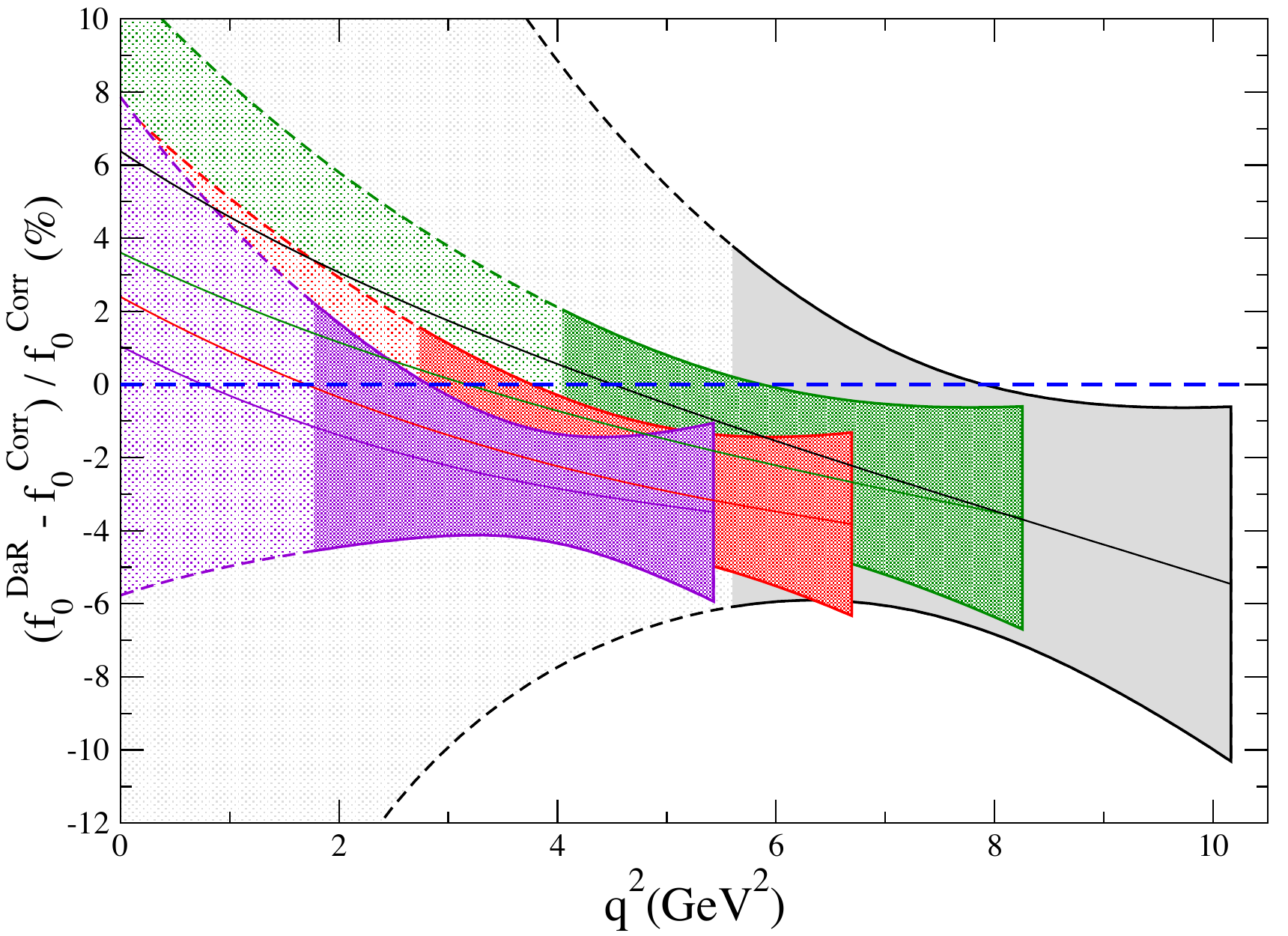}
}
\vspace{-0.75cm}
\caption{\it \footnotesize Percentage difference between the form factors extrapolated to the physical point obtained using either all available data (corrected for hypercubic effects) or using only data with the heavy meson at rest (uncorrected for hypercubic effects). Purple, red, green and grey bands correspond respectively to $m_h = 1.84, 2.13, 2.47$ and $2.87$ GeV. The short-dashed lines identify the extrapolation to values of $q^2$ not covered directly by the lattice data.}
\label{fig:HMaRcomparison_cont}
\end{figure}

We have performed the extrapolation to the physical pion point and to the continuum and infinite volume limits using two different fitting procedures: ~ i) including all available data taking into account the hypercubic corrections (see Eq.~(\ref{eq:vector_hypercubic})), and ~ ii) considering only the data with the heavy meson at rest neglecting hypercubic effects. In Fig.~\ref{fig:HMaRcomparison_cont} the (percentage) difference between the corresponding (correlated) results is shown and it indicates that the subtraction of hypercubic effects is required for obtaining properly the continuum limit.

\section{Conclusions}
\label{conclusions}

In this contribution we have presented the preliminary results of our study of the form factors describing the semileptonic transition between a heavy-light pseudoscalar meson, with masses above the physical D-meson one, and the pion.
Our analysis is based on the gauge configurations produced by ETMC with $N_f = 2 + 1 + 1$ flavors of dynamical quarks at three different values of the lattice spacing and with pion masses as small as $210$ MeV.
We have simulated a series of heavy-quark masses in the range $m_c^{phys} < m_h < 2m_c^{phys}$.

The Lorentz symmetry breaking due to hypercubic effects is clearly observed in the data as already found in Refs.~\cite{Lubicz:2017syv,Lubicz:2018rfs} in the case of the semileptonic form factors of the $D \to \pi(K) \ell \nu$ transitions.  The impact of the hypercubic contributions, included in the decomposition of the current matrix elements in terms of additional form factors, remains significant as the decaying meson mass increases and it should be taken into account for performing properly the continuum limit.

Therefore, we expect that hypercubic effects will play an important role in the lattice determination of the form factors governing the $B \to \pi \ell \nu$ semileptonic decays. 
Further investigations in this direction are in progress.

\bibliographystyle{JHEP}
\bibliography{lattice2018.bib}

\end{document}